\documentclass[prl,reprint,twocolumn,showpcs,amsmath,amssymb,superscriptaddress]{revtex4-1}
\usepackage{graphicx}
\usepackage{dcolumn}
\usepackage{bm}

\newcommand{\ket}[1]{$|#1\rangle$}
\usepackage{xcolor}

\begin{document}
\title{Electron-nuclear coherent spin oscillations probed by spin dependent recombination}

\author{S. Azaizia}
\author{H. Carr\`ere}
\affiliation{Universit\'e de Toulouse, INSA-CNRS-UPS, LPCNO, 135 avenue de Rangueil, 31077 Toulouse, France}
\author{J. C. Sandoval-Santana}
\author{V. G. Ibarra-Sierra}
\affiliation{Departamento de F\'isica, Universidad Aut\'onoma Metropolitana Iztapalapa, Av. San Rafael Atlixco 186, 
Col. Vicentina, 09340 Cuidad de M\'exico, M\'exico}
\author{V. K. Kalevich}
\author{E. L. Ivchenko}
\author{L. A. Bakaleinikov}
\affiliation{Ioffe Institute, 194021 St. Petersburg, Russia}
\author{X. Marie}
\author{T. Amand}
\affiliation{Universit\'e de Toulouse, INSA-CNRS-UPS, LPCNO, 135 avenue de Rangueil, 31077 Toulouse, France}
\author{A. Kunold}
\affiliation{\'Area de F\'isica Te\'orica y Materia Condensada, Universidad Aut\'onoma Metropolitana  Azcapotzalco, Av. 
San Pablo 180, Col. Reynosa-Tamaulipas, 02200 Cuidad de M\'exico, M\'exico}
\author{A. Balocchi}
\email{andrea.balocchi@insa-toulouse.fr}
\affiliation{Universit\'e de Toulouse, INSA-CNRS-UPS, LPCNO, 135 avenue de Rangueil, 31077 Toulouse, France}

\date{\today}
\begin{abstract}
We demonstrate the detection of coherent electron-nuclear spin oscillations related to the hyperfine interaction and revealed by the band-to-band photoluminescence (PL) in zero external magnetic field. 
On the base of a pump-probe PL experiment we measure, directly
in the temporal domain, the hyperfine constant of an electron coupled to a gallium 
defect in GaAsN by tracing the dynamical  behavior of the conduction electron spin-dependent 
recombination to  the defect site. The hyperfine constants and the relative 
abundance of the nuclei isotopes involved can be determined without the need of electron spin 
resonance technique and in the absence of any magnetic field. Information on the nuclear and electron spin relaxation
damping parameters can also be estimated from the oscillations damping and the long 
delay behavior.\end{abstract}
\maketitle
Electron and nuclear spins of well-isolated point defects in semiconductors are excellent candidates for understanding 
fundamental spin-coupling mechanisms or to model quantum information processing. 
The coupling through hyperfine interaction (HFI) represents
 a key spin mechanism in semiconductor systems: responsible for creating mixed electron-nuclear spin states, 
it has been shown to  be useful, e.g., for electron-nuclear spin transfer,  in controlling electron spin coherence time of  P donor sites in Si~\cite{kane_silicon-based_1998,Pla2012,PhysRevB.71.014401,Laucht2014,PhysRevB.74.195301,PhysRevB.92.115206} and the nitrogen-vacancy centers in diamond~\cite{PhysRevB.92.081301,zaiser_enhancing_2016,PhysRevB.93.035402,Sar2012,
PhysRevB.80.041201,PhysRevB.78.094303,PhysRevLett.113.246801}. 
The HFI is however also responsible for electron and nuclear spin relaxation and
 decoherence~\cite{RevModPhys.85.79}.
\begin{figure}[!h]
 \includegraphics[width=0.45\textwidth,keepaspectratio=true]{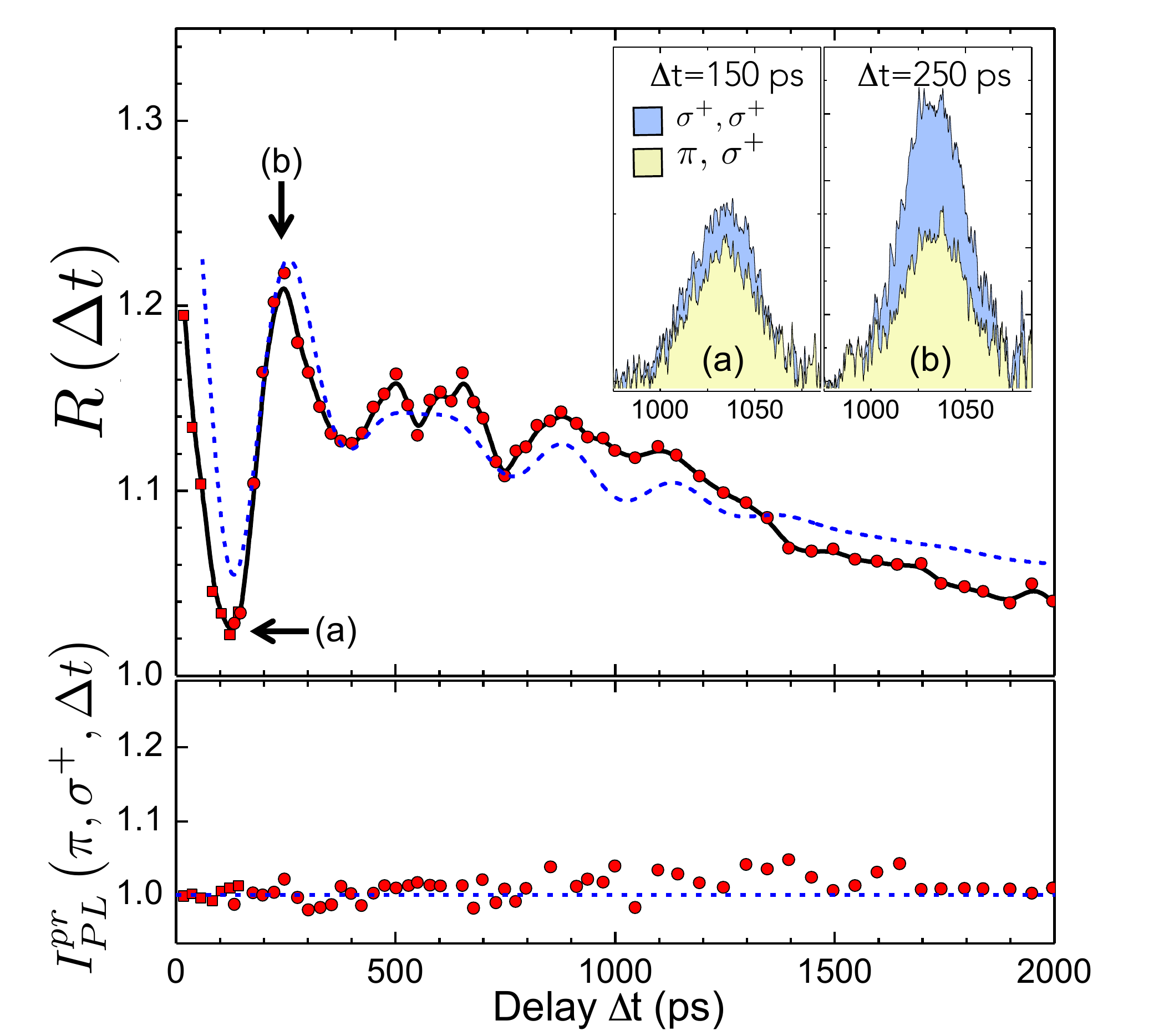}
 \caption{(Color online)  Top, symbols: The ratio of the probe pulse PL intensity 
 under a circularly to linearly polarized pump pulse $R(\Delta t$) measured as a function of the delay time $\Delta t$.
The time origin is  taken at the arrival of the pump pulse.
The black line is a guide to the eye, while the blue dashed line is a fit to the data 
according to Eq. (7). The inset 
presents the PL intensities of the circularly polarized probe pulse after a circularly (blue area) or linearly (yellow area) polarized pump pulse at two different delay times (a) and (b) indicated by the arrows in the main graph. Bottom, symbols: The normalized probe pulse PL intensity after a linearly polarized pump pulse excitation. The blue dashed line presents the results of the simulation setting the polarization of the pump pulse to linear. In both figures the red circles (squares) represent  the data measured
with the streak camera (photodiode) set-up.}
 \label{fig:figure_1}
\end{figure}
Similarly to nitrogen in diamond, on one side, and to shallow defects in silicon, on the other side,
 interstitial Ga$_{i}^{2+}$ defects in dilute nitride GaAsN~\cite{Kalevich2005,wang_room-temperature_2009} 
 unite the characteristics of deep and well isolated paramagnetic centers to an electrically 
 and optically addressable semiconducting system leading, e.g., to the giant spin-dependent photocondctivity effect~\cite{pssa_2007,zhao_2009,PhysRevB.83.165202}.
The incorporation of  nitrogen in (In)GaAs to form (In)GaAsN alloys gives
rise to paramagnetic interstitial centers composed of a Ga$_i^{2+}$ atom and a single 
resident electron~\cite{wang_room-temperature_2009}. 
These defect sites are at the origin of a very efficient  spin-dependent recombination 
 of conduction band (CB) electrons. This has proven, for instance, to 
be an effective tool for generating an exceptionally  high spin polarization (up to $\sim$100\%) of free and 
bound electrons in these nonmagnetic dilute nitrides semiconductors at room 
temperature~\cite{Kalevich2007}. 
The nuclear spin states of these defect ensemble has been shown to be accessible via a
measurement of the the circular polarization degree of band-to-band PL, while the defect nuclear spin
polarization in this model system can be tuned with different excitation parameters such as pump power, the
circular polarization degree of the incident light and through a weak external 
magnetic field~\cite{PhysRevB.85.035205,Kalevich2013,Ivchenko2016}.\\
Optically or electrically detected magnetic resonance techniques are consistently employed for manipulating and 
probing the defect spins through the hyperfine interaction, or again to identify the defect chemical nature and related 
spin dependent recombination pathways. Here, we demonstrate the experimental implementation of a new all-optical 
 detection scheme in zero external magnetic field based 
on a  PL pump-probe experiment  leading to the measurement, in the temporal domain, of the hyperfine 
constant of deep paramagnetic centers by directly tracing the hyperfine interaction dynamical features~\cite{Ibarra2016}. 
The hyperfine constants, defect configuration and the relative abundance of the isotopes involved can be 
determined without the need of electron spin resonance techniques and in the absence of any magnetic field. 
Information on the nuclear and electron spin relaxation damping parameters can also be estimated from the 
oscillations damping and the long temporal delay behavior.
\begin{figure}
 \includegraphics[width=0.45\textwidth, keepaspectratio=true]{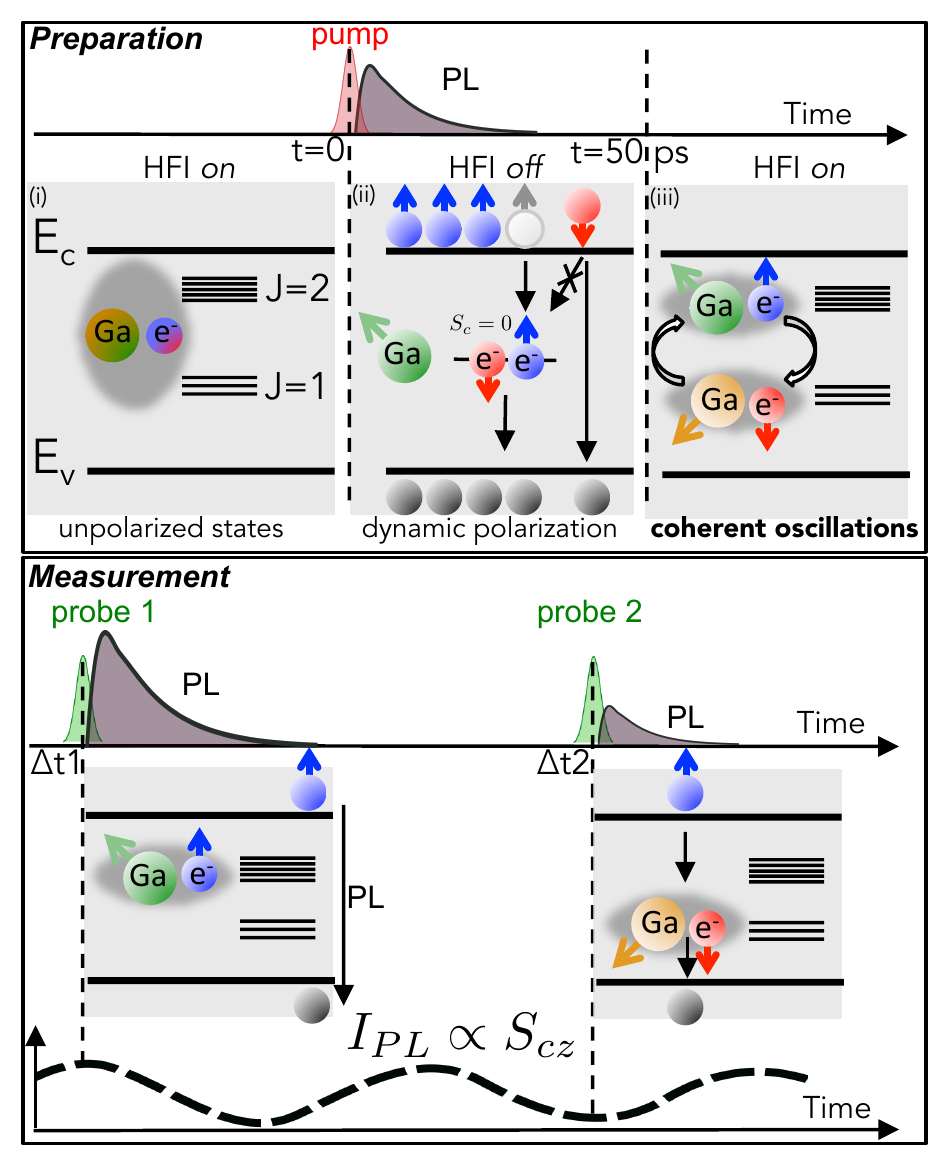} 
 \caption{(Color online) Schematic representation of the paramagnetic defect nuclear 
 and electron spin oscillations initiated by the pump circularly polarized pulse. The defect electron arrows indicate the orientation of the average spin $S_{cz}=\langle \hat{S}_{cz}\rangle$ projections. Top: preparation 
 of the coherent oscillations. Bottom: illustration of two extreme situations encountered by the probe beam:
 in the first situation (probe 1)  the majority of defect electrons have the same spin orientation as the CB ones, preventing the capture. The PL
 intensity is enhanced. In the second case (probe 2), the majority of defect electrons  and CB electrons have the opposite spin orientation, favoring capture: the conduction band is now depleted and the PL intensity is reduced.} 
 \label{fig:figure_2}
\end{figure}

The sample under study consists of a 100 nm thick GaAs$_{1-x}$N$_x$ epilayer (x=0.021)  grown by molecular beam epitaxy
 on a (001)  semi-insulating GaAs substrate  and capped with 10 nm GaAs.  The  sample has been investigated at 4 K by the optical orientation technique which relies on the successive transfer of the angular momentum of the exciting photons, using circularly
  polarized light, to the photogenerated electrons~\cite{Meier1984} and finally to the Ga$_i^{2+}$ nuclei. The excitation source is a mode-locked  Ti:Sapphire 
  laser emitting at 850 nm split into pump and probe pulses of equal intensity and 1.2 ps duration.
Although this configuration differs from a typical pump-probe experiments as the probe beam modifies 
on its turn the system similarly to the pump, this has proven to be the optimal experimental conditions to 
measure the HFI features. The sample is excited by focussing the pump and probe pulses to the same 50 $\mu$m 
diameter spot. The two pulses relative delay $\Delta t$ is controlled by an optical delay line and their polarizations 
independently set by a system of polarization optics. 
In order to evidence the HFI features, we have measured the PL intensity induced by a circularly
 polarized probe pulse as a function of   (i) the delay time between pump and probe pulses and
  (ii) the helicity of the pump pulse.  
In the following, we plot the influence of the pump pulse helicity on the probe pulse 
PL intensity $I_{PL}^{\mathrm{pr}}$ by computing the ratio  
\begin{equation} \label{Rdelta}
R(\Delta t)= \frac{I_{PL}^{\mathrm{pr}}(\sigma^+,\sigma^+,\Delta t)}{I_{PL}^{\mathrm{pr}}(\pi,\sigma^+,\Delta t)}\:,
\end{equation}
where the notation ($\sigma^+/\pi$ , $\sigma^+$, $\Delta t$) indicates, in arrival order,  
the polarization of the pump pulse, probe pulse and their respective delay.
For the measurements at short delays ($\Delta t<$100 ps for which the PL intensities 
of the two pulses may partially overlap),  the intensity of the probe pulse has been modulated by a mechanical
chopper and the PL intensity of the 
probe pulse has been measured with a photodiode 
connected to a lock-in amplifier. 
For greater delays, a S1 photocathode streak camera coupled to an imaging spectrometer has been employed.
Figure~\ref{fig:figure_1} (top, symbols) presents $R(\Delta t$) measured for an excitation power 
$P_{exc}$=4 mW. The trace presents an oscillating behaviour showing that 
$I_{PL}^{\mathrm{pr}}$ can be periodically amplified by adjusting the delay 
time $\Delta t$ when the pump pulse is circularly polarized. For longer delay times the oscillations lose 
visibility and a monotonous decrease of the amplifying  effect is observed. 
Insets in Figure~\ref{fig:figure_1} display the recorded $I_{PL}^{\mathrm{pr}}$ 
data at two extreme points of the oscillations where the probe PL intensity is strongly 
increased ($\Delta t$=250 ps) or only weakly modified  ($\Delta t$=150 ps). Figure~\ref{fig:figure_1}, (bottom, symbols)
presents the PL intensity of the probe pulse recorded after a linearly polarized pump pulse. In this case
no significant variation of the PL intensity is measured.
Below we will show that the oscillating and decay features of 
$R(\Delta t)$ can be directly linked to the coherent oscillations of the electron-gallium system
induced by the hyperfine-interaction.

The principle of the observation is described in terms of the 
model depicted in  Figure~\ref{fig:figure_2}  and based on the following considerations. 
The HFI hamiltonian $\mathcal{\hat{H}}_{HFI}=\mathcal{A} \boldsymbol{\hat{I}}\cdot \boldsymbol{\hat{S}}_c$ 
 of gallium defects with spin $I$=3/2 and a singly trapped electron  of spin $S_c$=1/2 
 (where $\mathcal{A}$  is the hyperfine interaction constant) leads to split triplet-quintet eigenstates at zero 
 magnetic field \ket{J,M}=$\sum_{s,m}C_{s,m,M}^{\frac{1}{2},\frac{3}{2},J}$ \ket{s,m}, where $C_{s,m,M}^{\frac{1}{2},\frac{3}{2},J}$  are the Clebsch-Gordan coefficients ($m=\pm$ 
  1/2, $\pm$ 3/2, $s=\pm$1/2 are the 
  nuclear and electron spin projections on the $z$ axis,  and $M = - J,\dots,J$) with total spin $J=1,2$. Before illumination by the  pump pulse, the
   eight hyperfine states are statistically equi-populated in our experimental conditions. The preparation of the defect spin polarisation
    by the pump pulse proceeds as follows: A left-handed circularly polarized pumping above the 
    band gap creates preferentially spin-up conduction band electrons (holes quickly lose their spin orientation, and 
are considered unpolarized~\cite{PhysRevLett.89.146601}). 
The conduction band electrons are very rapidly captured by the gallium interstitial defects ($\tau_{e}\lesssim$ 10 ps)
 forming a two-electron spin singlet~\cite{kalevich_hanle_2009, wang_room-temperature_2009}: The hyperfine coupling is now {\it off}.
It follows a fast recombination of one of the two defect electrons with an unpolarized hole.
Due to the photogeneration of a CB electron spin polarization, the spin dependent recombination 
statistically drives the defect electrons to the same average spin orientation as the conduction 
band electrons~\cite{Kalevich2005,PhysRevB.90.115205}.
The  recombination of one of the center paired electron with an unpolarized hole  is fast, typically occurring on 
a time scale $\tau_{h}\sim$ 30 ps~\cite{Kalevich2005,pssa_2007,wang_room-temperature_2009}, and the HFI is re-enacted.
At the re-establishment of the HFI the remaining electron spin state is projected onto the total spin 
eigenstates \ket{J,M}, leaving  the defect system in a superposition of states between the $J=1$ and $J=2$ 
 hyperfine levels as $\tau_{h}\cdot 2\mathcal{A}/\hbar \ll$1. Second, the time scale of the recombination 
 ensures as well a relatively constant phase among the  ensemble of the defect centers.  
{\it At this point, the quantum system periodically oscillates between 
 the $J=1$ and $J=2$  states which results in $S_{cz}$ oscillations between the $\pm1/2$ states}.
Being the hyperfine interaction energy $2 \mathcal{A}\sim$15 $\mu$eV~\cite{buya_2009},  the defect 
preparation time is sizeably shorter than the oscillation period $T=h/2\mathcal{A}\sim$ 250 ps.
The probe beam (Figure~\ref{fig:figure_2}, bottom panel) can now encounter  two extreme situations.
In the first case (probe 1) the majority of defect electrons have the same average spin orientation as the 
CB ones, preventing the capture. The PL  intensity is enhanced. In the second case 
(probe 2), the majority of defect electrons have the opposite average spin orientation than the CB electrons, favoring
the capture: the PL is now reduced as the conduction band is depleted.
In principle this process would result in a partially oscillating PL intensity in 
a spin polarized system with sufficiently long PL characteristic decay time $\tau_{PL}$ in a single pulse experiment. 
Here, being the PL decay time shorter than the oscillation period T,  a pump-probe PL technique,  employing a second pulse, samples in time the
coherent oscillation of the hyperfine system. It is important to note that longer PL decay times $\tau_{PL}$
 (as can be obtained by higher power excitation in this systems~\cite{JOP}) are here not desirable as this will smear the
initialization of the oscillation leading to faster dephasing.
 If however the pump pulse is linearly polarized, no dynamical polarization can occur and the probe pulse 
spin dependent recombination will be insensitive to the delay time.
A key feature is that this all-optical approach does not require any external magnetic field which
might modify the spin relaxation damping parameters.
 Let us now turn our attention to the kinetics of  photoelectrons excited by the probe pulse in the conduction band.
We can get a qualitative and analytical understanding of the oscillating behavior of $R(\Delta t)$ according 
to the following argument. Neglecting the electron spin relaxation, 
the CB electron rate equations can be described by
\begin{eqnarray} \label{dynamics}
&&\frac{dn_+}{dt}+2c_n n_+N_- + \gamma_r n_+p=0\:,\\
&&\frac{dn_-}{dt}+2c_n n_-N_+ + \gamma_r n_-p =0\:, \nonumber
\end{eqnarray}
where $n_{\pm}(t)$ are the densities of the conduction photoelectrons with spin up ($+$) and down ($-$) 
excited by the probe pulse arriving with the time delay $\Delta t$, $c_n$ is the constant of the conduction-electron 
trapping rate by paramagnetic centers, $p$ is the hole concentration (due to rapid spin relaxation the 
holes are unpolarized), $\gamma_r$ is the bimolecular recombination constant, and $N_{\pm}$ are the
 concentrations of single-electron defects with the electron spin $\pm 1/2$ with $N_++N_-=N_1$. Due to the prior pump pulse 
$N_+$ and $N_-$ are different if the pump is circularly polarized and coincide for the linearly polarized
pump excitation. Since (i) the electron capture is much more effective as compared to the interband 
recombination and  (ii) $N_{\pm}$ vary slowly within the capture times $(c_n N_{\pm})^{-1}$, the time
dependence of $n_{\pm}$ is described by $n^{\rm pr}_{\pm}\exp{[- 2c_n N_{\mp}(\Delta t) t]}$, 
where $n^{\rm pr}_{\pm}$ are the electron densities injected by the probe pulse. For a sufficiently 
weak photoexcitation the measured ratio (\ref{Rdelta}) is described by
\begin{equation} \label{Rdelta2}
R(\Delta t) - 1 \propto \left(n^{\rm pr}_+ - n^{\rm pr}_- \right) \left[N_+(\Delta t) -  N_-(\Delta t)\right] \:.
\end{equation}
For circularly polarized pump pulses the values $N_{\pm}(\Delta t)$ consist of the oscillating 
and non-oscillating parts
\begin{equation} \label{NDelta}
N_{\pm}(\Delta t) = N_{\pm,0}\pm\frac{\delta N}{2}\cos {(\Omega \Delta t)}\:,
\end{equation}
where $\hbar\Omega=2 \mathcal{A}$ is the hyperfine splitting between the electron-nuclear spin quintet 
and triplet with the angular momenta $J=2$ and $J=1$, respectively.

The oscillating time behaviour of $N_{\pm}(\Delta t)$ can be understood in terms of the 
spin-density-matrix approach. In equilibrium the spin density of single-electron defects, 
$\rho_{J',M';J,M}$, is diagonal with equally populated 
sublevels: $\rho_{J',M';J,M} = (N_1/8)\delta_{J'J} \delta_{M'M}$. The pump pulse generates 
CB photoelectrons with densities  $n^{\rm pm}_{\pm}$ which are immediately captured by 
single-electron defects according to Eqs.~(\ref{dynamics}) and form the electron pair states 
with density $N_2 = n^{\rm pm}_+ + n^{\rm pm}_-$. The remaining single-electron defects 
acquire spin polarization. Immediately after the pulse, i.e. at 
$\Delta t = 0$, one has $N_{\pm}(0) = N_c/2 - n_{\mp}^{\rm pm}$, 
where $N_c=N_1+N_2$ is the total density of the deep paramagnetic centers. This equation can be rewritten as
\[
N_{\pm}(0) = \sum\limits_{m} \rho_{\pm \frac12,m;\pm \frac12,m}(0) 
\]
in terms the spin-density matrix $ \rho_{s',m';s,m}(0) = \delta_{s's} \delta_{m'm} \left( N_c/8 - n^{\rm pm}_{-s} \right)$
taken in the basis $|s,m\rangle$. In the basis $|J,M\rangle$ this equation is rewritten as
\begin{equation} \label{rho0}
N_{\pm}(0) = \sum\limits_{mJ'J}  D_{J',J;\pm 1/2, m} \rho_{J',m \pm \frac12;J,m \pm \frac12}(0) \:,
\end{equation}
where
\[
D_{J',J;s,m} = C^{\frac12 \frac32 J'}_{s, m, s+m} C^{\frac12 \frac32 J}_{s, m, s+m}
\]
and the Clebsch-Gordan coefficients $C^{\frac12 \frac32 J}_{s m M}$  relate the standard bases
 $|s,m\rangle$ and $|J,M\rangle$. The components $\rho_{J',M;J;M}(0)$ can be readily expressed 
 via $N_c, n^{\rm pm}_+$ and $n^{\rm pm}_-$, among them there are those with $J'=J$ and $J' \neq J$.
Neglecting the spin relaxation we have
\[
\rho_{J',M;J,M} (\Delta t) = \rho_{J',M;J,M}(0) {\rm e}^{- {\rm i} \Omega (J'-J)\Delta t}\:.
\] 
Therefore, the oscillating part of Eq.~(\ref{NDelta}) is contributed from the off-diagonal 
spin-matrix components with $J' \neq J$. A straightforward calculation gives
\[
 N_{+,0} -  N_{-,0} = \frac38 \left( n_+^{\rm pm} - n_-^{\rm pm}\right)\:,\:
\delta N= \frac58 \left( n_+^{\rm pm} - n_-^{\rm pm}\right)\:. 
\]
The last terms in the left-hand sides of Eqs.~(\ref{dynamics}) describe the radiative recombination 
and the PL intensity. Retaining all the factors we can present Eq.~(\ref{Rdelta2}) in the final form
\begin{equation}
R(\Delta t) - 1 = \alpha+ \beta \cos\left(\Omega \Delta t\right)\:,
\label{eq:sdr_1}
\end{equation}
where the coefficients $\alpha$ and $\beta$ are respectively given by $P_i \left( N_{+,0} - N_{-,0}\right)/N_c $ and $P_i \delta N/ N_c$ 
with $P_i$ being the initial degree of the pump-induced spin polarization $ \left( n_+^{\rm pm} - n_-^{\rm pm}\right)/ \left( n_+^{\rm pm} + n_-^{\rm pm}\right)$. 
In the presence of two isotopes, as is the case for gallium atoms, the cosine function in Eq.~(\ref{eq:sdr_1}) should be 
replaced by a sum of two cosine functions with the frequencies $\Omega_1, \Omega_2$ and relative abundances $f_1,f_2$. 
Allowance for the spin relaxation results in a multiplication of $\alpha$ by $\exp{(- \Delta t/\tau_{sc})}$ and of
 $\beta$ by  $\exp{(- \Delta t/T_2)}$, where $\tau_{sc}$ is the bound-electron spin relaxation time and $T_2$
 is the decoherence time of electron-nuclear spin oscillation which is affected by both homogeneous 
relaxation processes and inhomogeneous broadening of the hyperfine splitting. Thus the experimentally
determined $R(\Delta t)$ can be quantitatively compared with
\begin{eqnarray}
&&R(\Delta t) - 1 = \alpha {\rm e}^{- \Delta t/\tau_{sc}} \\&&\mbox{}\hspace{1 cm}+
\:
 \nonumber\beta {\rm e}^{- \Delta t/T_2} \left[ f_1 \cos\left(\Omega_1 \Delta t\right) + f_2 \cos\left(\Omega_2 \Delta t\right) \right]\:, 
\label{eq:sdr2}
\end{eqnarray}
where the positive values $f_1$ and $f_2$ are normalized by the condition  $f_1 + f_2 = 1$.
According to the  mechanism described, the  circularly polarised probe pulse will sample 
the oscillating behaviour of the hyperfine coupling instantaneously and at different delays.
We obtain a modulation of the probe beam PL intensity directly tracing the hyperfine interaction in the time domain.
In the case of gallium atoms, the two stable isotopes  $^{69}$Ga and $^{71}$Ga have the relative abundances  $f_1$=0.6018 and $f_2$=0.3982 respectively, 
and their hyperfine constants differs by a factor $\mathcal{A}_2/\mathcal{A}_1$=1.27.
The hyperfine constant (Table I~\cite{buya_2009}) will however depend on the particular defect location which is
determined mainly by growth and annealing conditions. 
\begin{figure}
 \includegraphics[width=0.45\textwidth, keepaspectratio=true]{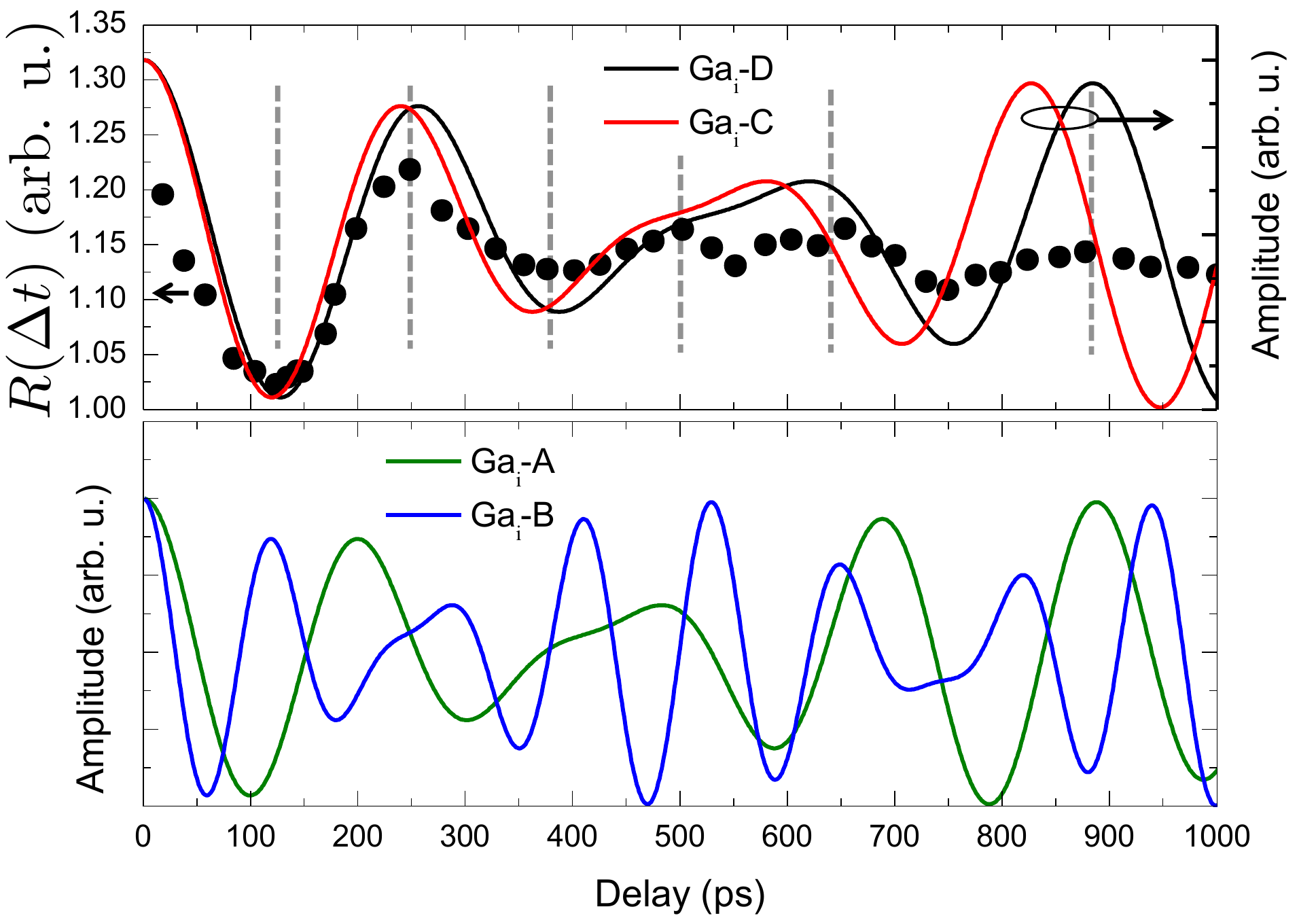} 
 \caption{(Color online) Detail of the experimental data for short delays (circles) superimposed to
 simple cosine beat patterns (solid lines) at the frequencies of the different interstitial configurations. The vertical dashed lines indicate the position of the extrema of the experimental data.} 
 \label{fig:figure_3}
\end{figure}
Figure 3 details $R(\Delta t)$ for short delays superposed to a simple  beating pattern
composed of a sum of simple cosines functions at the frequencies of the different gallium interstitial locations as reported in Table 1. The experimental data present a first
destructive interference pattern in the 500 ps to 600 ps range. This rules out the 
occurrence of  Ga$_i$-A and Ga$_i$-B as their higher oscillating frequencies are incompatible with the 
experimental observation. Ga$_i$-C and Ga$_i$-D present very similar hyperfine constant which causes the beating pattern to dephase slightly. Nevertheless, the experimental data presented in Figure 3 (top) allow us to unambiguously identify  Ga$_i$-D as the dominant interstitial configuration.\\
The dashed line in Figure 1 (top) represents a fit to the data using Eq. (7),  the hyperfine constants of gallium D interstitial defects and setting the pump pulse polarization to circular. The best results are obtained with $\tau_{sc}$=1700 ps, $T_2$=350 ps. 
This proves that a precise determination of the defect nature and configuration
can be obtained by this PL pump and probe scheme.
The mechanisms responsible for the coherence loss can be multiple. First of all, the measurement maps the coherent oscillations of the ensemble of Ga centers present under the 
excitation spot whose intensity strongly varies from the excitation spot center to the edge.
 Second, the HFI sampling cannot be considered as strictly instantaneous but averaged over the CB electron lifetime $\tau_{PL}$, allowing us only to infer a minimum limit for the coherence time decay. Third, the stochastic nature of the the trapping and recombination can also be ascribed as a possible source of coherence loss. Finally, there might be some inhomogeneities from center to center
 resulting in slight fluctuations of the hyperfine coupling constant.
\begin{table}
\begin{tabular}{lcccc}
\hline
\hline
\\
Ga$_i$ location & A & B & C & D\\
\\
\hline
\\
$\mathcal{A}_1$ ($^{69}$Ga) ($\times$10$^{-4}$ cm$^{-1}$) & 745 & 1230 & 620 & 580\\
$\mathcal{A}_2$ ($^{71}$Ga) ($\times$10$^{-4}$ cm$^{-1}$) & 968.5 & 1562 & 787.4 & 736.6\\
\\
\hline
\hline
\end{tabular}
\caption{The hyperfine interaction constants for the two naturally stable isotopes of gallium
 in the four different interstitial configurations occurring in dilute nitrides (In)GaAsN~\cite{buya_2009}.}  
\end{table}
 The dotted line in Figure 1 (bottom) reports the results of the simulation setting the first pulse
 polarization to linear. As expected from the previous discussion, $N_+$ and $N_-$ are now identical and no oscillations are observed. 
 
In conclusion, we have demonstrated a possibility of measuring the electron-nuclear spin oscillations related to the hyperfine interaction in dilute nitride semiconductor 
paramagnetic centers by monitoring the band to band PL in the absence of any magnetic field.  
The hyperfine constants and the relative abundances of the nuclei involved can be unambiguously 
determined without the need of electron spin resonance techniques. Information on the nuclear and electron spin relaxation
damping parameters of the paramagnetic center can also be estimated from the oscillations damping and the long time delay behavior.
This zero magnetic field detection scheme based solely on the spin dependent recombination should be applicable to other materials besides GaAsN, like 2D crystals as, for instance, paramagnetic centers in h-BN~\cite{Trang_2016}.

\acknowledgments
We acknowledge funding from LIA CNRS-Ioffe RAS ILNACS. E.L.I. thanks the RFBR (Grant No. 17-02-00383). 
V.K.K. acknowledges the financial support of the Government of Russia (Project No. 14.Z50.31.0021).  A.K. gratefully 
appreciates the financial support of ÒDepartamento de Ciencias B\`asicas UAM-AÓ grant numbers 2232214 
and 2232215. J.C.S.S. and V.G.I.S. would like to acknowledge the support received from the ÒBecas de 
Posgrado UAMÓ scholarship numbers 2151800745 and 2112800069. X.M. also thanks Institut Universitaire de France.
This work was supported by Programme Investissements d'Avenir under the program ANR-11-IDEX-0002-02, reference ANR-10-LABX-0037-NEXT.
\bibliography{biblio}
\end{document}